\documentclass{article}
\usepackage{amsmath}

\begin{document}
\vskip 1truein
\begin{center}
{\Large {\bf ON CONSISTENCE OF MATERIAL COUPLING IN \\ A GL(3,R)
GAUGE
 FORMULATION OF GRAVITY }} \vskip 10pt {\bf ROLANDO
GAITAN}${}^{a, }${\footnote {e-mail:
rgaitan@fisica.ciens.ucv.ve}} \\
${}^a${\it Grupo de F\'{\i}sica Te\'orica, Departamento de F\'\i
sica, Facultad de Ciencias y Tecnolog\'\i a, \\
Universidad de
Carabobo,A.P. 129
Valencia 2001, Edo. Carabobo, Venezuela.}\\
\end{center}
\vskip .2truein
\begin{center}
\begin{minipage}{4.5in}\footnotesize\baselineskip=10pt
    \parindent=0pt

A covariant scheme for material coupling with $GL(N,R)$ gauge
formulation of gravity is studied. We revisit a known idea of a
Yang-Mills type construction, where quadratical power of
cosmological constant have to be considered in consistence with
vacuum Einstein's gravity. Then, matter coupling with gravity is
introduced and some constraints on fields and background appear.
Finally, exploring the $N=3$ case we elucidate that introduction
of auxiliary fields decreases the number of these constraints.

\end{minipage}
\end{center}

\vskip .2truein
\begin{center}
\begin{minipage}{4.5in}\footnotesize\baselineskip=10pt
    \parindent=0pt



\end{minipage}
\end{center}
\vskip .2truein

The old problem about construction of gauge theory for gravitation
cover a large list of approaches (among others, see some
fundamental references$^{1-9}$). Starting with Utiyama$^{1}$ who
was the first to recognize the ''gauge'' character of
gravitational field, passing by constructions which relax symmetry
property of connection (i.e., Riemann-Cartan spacetime) and others
where  metric-compatible condition is also removed, arising
theories based on non-Riemannian geometries$^{8,9}$.

 However, in order to compare and to explore possible
inconsistencies between solutions coming from Einstein's theory
and those of a gauge formulation of gravity, here we focus our
attention in a construction based on a subgroup of the
affine$^{8,9}$, this means $GL(N,R)$ as gauge group$^{5}$.
Obviously, choosing this subgroup causes limitations related to
the context of supersymmetric theories which demand translation
symmetries, as an example. But having in mind to show
inconsistencies, it is sufficient the aforementioned construction
in a Riemannian spacetime.

It is well known that a $GL(N,R)$ Yang-Mills type lagrangian
density will
 be related to a purely quadratic lagrangian density on
Riemann-Christoffel curvature. These lagrangians take great
interest, because, from a point of view of standard field theory
these yield theories where the renormalization problems are much
less severe$^{10}$; from the string theory point of view, this
class of terms does appear at low energy limit of effective
lagrangian for gravity$^{11}$, among other things.

The aim of this letter is to explore a general and covariant
scheme for non minimal coupling of material fields in $N$
dimension with $GL(N,R)$ connection. So, we study consistence
between solutions coming from this type of theory and the
Einstein's gravity ones. At the trivial vacuum case, last
requirement means to consider quadratical power of cosmological
constant.

On the other hand, when material fields are turned on, it is
observed that consistence gives rise to constraints on these
fields and background (i.e., cosmological constant). The last step
consists of including action terms dependent on auxiliary fields,
showing that some restrictions can be avoided. As a useful scenery
to explore and illustrate coupling we will study the particular
case in $2+1$ dimension.

 A brief review about notation is presented. Let
${V_{\mu}}^a$ be the components of a tensorial object with curved
and lorentz indexes, defined in an $N$ dimensional space provided
with metric $g_{\mu \nu }$, curved coordinates
 $x^\mu $, with $\mu ,\nu ,...=0,1,...,N-1$ and locally flat $\xi ^a $,
  with $a,b,c,...=0,1,...N-1$
(the Minkowski metric is $\eta _{ab} = diag(-1,+1,...,+1)$). Then,
introducing the spin (i.e., ${\omega _{\mu b}}^a $) and  affine
(i.e., ${\Gamma ^\lambda }_{\mu \nu }$) connections, the well
known covariant derivative is $ D_\mu {V_\nu }^a = \partial _\mu
{V_\nu }^a  +{\omega _{\mu b}}^a {V_\nu }^b - {\Gamma ^\lambda
}_{\mu \nu } {V_\lambda }^a $\,, etc. Particularly,   property
$D_\mu {e_\nu }^a =0$ is taken, where ${e_\nu }^a$ is the {\it
n-bein }which satisfies $g_{\mu \nu } = {e_\mu }^a {e_\nu }^b \eta
_{ab}$. From this, follows that the torsion can be written in the
form
\begin{equation}
{T^\lambda }_{\mu \nu } \equiv {\Gamma ^\lambda }_{\mu \nu } -
{\Gamma ^\lambda }_{\nu \mu } = {e^\lambda }_a (\partial_\mu
{e_\nu }^a - \partial_\nu {e_\mu }^a + {\omega _{\mu \nu }}^a -
{\omega _{\nu \mu }}^a)
 \, \, , \label{eq1}
\end{equation}
where, ${\omega _{\mu \nu }}^a \equiv {e_\nu }^b \,{\omega _{\mu b
}}^a$, etc.

If the matrix elements for $GL(N,R)$ and lorentz transformations
are defined as ${(U)^\alpha }_\mu \equiv \frac{\partial {x^\prime
}^\alpha }{\partial x^\mu}$ and ${(L)^a}_b \equiv \frac{\partial
{\xi^\prime }^a }{\partial \xi^b}$, covariant behavior of
derivative $D_{\mu}$ demands following transformation rules for
connections
\begin{equation}
{\omega ^\prime }_\mu  =L \omega _\mu L^{-1} + L \partial _\mu
L^{-1} \, \, , \label{eq2}
\end{equation}
\begin{equation}
{A_a}^\prime =UA_a U^{-1} + U \partial _a U^{-1} \, \, ,
\label{eq3}
\end{equation}
where we have introduced notation
\begin{equation}
{(A_a)^\mu }_\nu \equiv {e^\alpha }_a {\Gamma ^\mu }_{\alpha \nu }
 \, \, . \label{eq4}
\end{equation}
It can be observed that the object (4) is a $GL(N,R)$ connection
which transforms like a lorentzian vector in flat index. So,
$GL(N,R)$ is chosen as the structure group, and will be assumed
that the fibre projection, the transition functions, etc. are
given.

The Riemann-Christoffel curvature tensor (i.e.,
${R^\alpha}_{\sigma \mu \nu }$) can be constructed through
application of commutator $[D_\mu ,D_\nu]\equiv D_\mu D_\nu -
D_\nu D_\mu $ on a rank one tensor, in other words $[D_\mu
,D_\nu]V^\alpha ={R^{\alpha}}_{\sigma \nu \mu } V^\sigma -
{T^\lambda }_{\mu \nu } D_\lambda V^\alpha $\,, for all $V^\alpha
$ implies ${R^\sigma}_{\alpha \mu \nu } = \partial _\nu {\Gamma
^\sigma }_{\alpha \mu  } - \partial _\mu {\Gamma ^\sigma }_{\alpha
\nu } + {\Gamma ^\lambda }_{ \alpha \mu} {\Gamma ^\sigma
}_{\lambda \nu }
 -  {\Gamma ^\lambda }_{\alpha \nu} {\Gamma ^\sigma }_{\lambda \mu  }$. Using this definition,
the Riemann-Christoffel tensor components can be written in the
form
\begin{equation}
{R^\sigma}_{\alpha \mu \nu }=  {e_\mu}^a {e_\nu}^b{(F_{ab})^\sigma
}_\alpha
 \, \,, \label{eq5}
\end{equation}
where
\begin{equation}
{(F_{ab})^\sigma }_\alpha \equiv {\big( D_b A_a - D_a A_b + [A_a ,
A_b ]+ {T^c }_{a b} A_c -{T^c }_{a b} T_c \big)^\sigma }_\alpha\,
\,, \label{eq6}
\end{equation}
is a Yang-Mills like curvature with torsion contribution. Notation
means ${(T_c)^\sigma }_\alpha \equiv e_{\mu c}g^{\sigma \nu}
{T^{\mu }}_{\nu \alpha}$\,, etc.

In order to study the relationship between solutions obtained from
a Yang-Mills type lagrangian formulation defined
 with curvature $F_{ab}$, and solutions of Einstein's gravity
with cosmological constant in a Riemannian space ($EG\lambda$), we
will consider that Ricci tensor satisfies the field equation of
Einstein (i.e., $ R^{\alpha \beta }-\frac{g^{\alpha \beta
}}2R-\lambda g^{\alpha \beta}=-8\pi G T^{\alpha \beta }$, where
$T^{\alpha \beta }$ is the energy-momentum tensor associated to
material fields and $\lambda$ is the cosmological constant). This
ends the review.

We proceed noting the formulation without matter. Let the gauge
invariant action for $T^{\alpha \beta } =0$ be
\begin{equation}
S_o = \int d^N x \sqrt{-g} \,\, (-\frac 14 \, tr\, F^{ab}F_{ab} +
\Lambda ) \,\, , \label{eq7}
\end{equation}
where $\Lambda $ would be related to cosmological constant.
Observe that action (7) matches with a standard
Riemann-Christoffel quadratic lagrangian theory in a torsionless
space-time. In dynamical analysis we will assume a Palatini's
variational principle type (i.e., variations on $GL(N,R)$
connection and {\it n-bein}
 ) thinking about a general case where ${T^\lambda }_{\mu \nu }\neq
 0$. Afterwards, one can evaluate field equations in a particular
 space-time. An alternative starting point would be to consider a
 standard lagrange multiplier method, where the torsionless
 condition works as constraint on connection, and can be obtained
 the same physical results that we show in next.

Functional variation on connection $A$ gives
\begin{equation}
\delta _A S_o=\int d^N x \sqrt{-g}\,\,tr\, E^\sigma \delta
A_\sigma  \,\, , \label{eq8}
\end{equation}
up to a boundary term. Notation means ${(
 E^\sigma )^\lambda
}_\alpha \equiv {( \frac 1{\sqrt{-g}}\,\,\partial _\mu
(\sqrt{-g}\,\,F^{\lambda \mu }) + [F^{\mu \lambda  } ,A_\mu -T_\mu
])^{\sigma} }_\alpha $\,, and in a torsionless space-time must be
${(E^\sigma )^\lambda }_\alpha = D_\mu {(F^{\lambda \mu})^{\sigma}
}_\alpha $ (i.e., field equation is $D_\mu F^{\lambda \mu}=0$).
This last relation can be written in terms of Ricci tensor with
the help of Bianchi's identities in the form
\begin{equation}
 (E_\sigma )_{\lambda \alpha } = D_\alpha R_{\lambda \sigma } - D_\sigma R_{\lambda \alpha }
\,\,. \label{eq9}
\end{equation}
From this, the torsionless solutions for $\delta _A S_o=0$ are
those
 of de Sitter and Anti de Sitter (dS/AdS) type.
Then, taking the trivial solution
 $R_{\alpha \beta }=-(2\lambda /(N-2))
g_{\alpha \beta}$ (unique solution under this conditions), the
curvature $F_{ab}$ can be evaluated using eq.(5):
\begin{equation}
{(F_{ab})^\alpha }_\mu = \frac{2\lambda}{(N-2)(N-1)} ({e^\alpha
}_b e_{\mu a}-{e^\alpha }_a e_{\mu b})
 \, \,. \label{eq10}
\end{equation}

Next we consider {\it n-bein}'s variation on action $S_o$,
obtaining the field equation
\begin{eqnarray}
\,tr\,F_{db}{F_c}^b - \frac 14 \,\eta _{dc}\,tr\,F^{ab}F_{ab} +
\Lambda \,\eta _{dc}=0 \,\, , \label{eq11}
\end{eqnarray}
and demanding consistence with eq.(10) it says that
\begin{eqnarray}
\Lambda =-\frac{2(N-4)}{(N-2)^2(N-1)}\lambda ^2 \,\, .
\label{eq12}
\end{eqnarray}
In other words, this condition guarantees that dS/AdS are trivial
solutions for extremal of action (7). It can be noted  that
constant $\Lambda$ does not distinguish between dS or AdS in this
lagrangian formulation in contrast with the Hilbert-Einstein's
one. However, $\Lambda$ establishes other type of classification.
When $N=3$, $\Lambda $ takes the value  $\lambda ^2 \geq 0$, while
cosmological constant does not explicitly appear in the action for
$N=4$. On the other hand, if $N > 4$ one have $\Lambda \leq 0$.
Immediately one can say that the physical content of these classes
of $\Lambda$ are related to the shift on the Hamiltonian.

The model for coupling with matter through the connection $A_a$ is
outlined. We explore a possible general covariant scheme for non
minimal coupling without auxiliary fields in the following manner
\begin{equation}
S=S_o + \int d^N x \sqrt{-g}\,\big(\ell(e,\psi) + 4\pi
G\,\,tr\,M^{ab}(e,\psi)F_{ab} \big)
 \, \,, \label{eq13}
\end{equation}
where gauge invariant functional $\ell(e,\psi)$ and tensor
$M^{ab}(e,\psi)$ are functionals on {\it n-bein} and material
fields. From these definitions, it follows that, after an
integration of action (13) it can be obtained ``current'' (i.e.,
minimal coupling) and  ``mass'' (i.e., Proca type coupling) terms.
Now, our problem is to explore the shape of objects $\ell(e,\psi)$
and $M^{ab}(e,\psi)$ requiring consistence with Einstein's
solutions.

Here, we propose
\begin{equation}
{(M^{\alpha \beta})^\mu }_\nu = {(N^{\alpha \beta \sigma
\rho})^\mu}_\nu \,\,T_{\sigma \rho} + {(n^{\alpha \beta})^\mu}_\nu
\, \,, \label{eq14}
\end{equation}
where objects $N^{\alpha \beta \sigma \rho}$ and $n^{\alpha
\beta}$ depend only on metric. Their general form, in consistence
with symmetry properties are
\begin{eqnarray}
{(N^{\alpha \beta \sigma \rho})^\mu}_\nu \equiv c_1 \big(g^{\mu
\alpha } {\delta ^\beta }_\nu g^{\sigma \rho} -g^{\mu
\beta}{\delta ^\alpha }_\nu g^{\sigma \rho}\big)
 + c_2\big(g^{\rho \alpha } {\delta ^\beta }_\nu g^{\sigma \mu}-g^{\rho \beta}{\delta
^\alpha }_\nu g^{\sigma \mu}\big) \nonumber \\+ \,c_3\big(g^{ \mu
\alpha } g^{\rho \beta }{\delta ^\sigma }_\nu -g^{\mu
\beta}g^{\rho \alpha }{\delta ^\sigma }_\nu \big)
 \, \,, \label{eq15}
\end{eqnarray}
\begin{equation}
{(n^{\alpha \beta})^\mu}_\nu \equiv a \big(g^{\mu \alpha } {\delta
^\beta }_\nu - g^{\mu \beta}{\delta ^\alpha }_\nu \big)\, \,,
\label{eq16}
\end{equation}
and $c_1$, $c_2$, $c_3$, $a$  are real free parameters.

It is very important to note that, as an illustration in this
first approach to a coupling scheme we consider a system whose
energy-momentum tensor does not explicitly depend on connection,
just for simplicity. One can think about a class of this type of
systems. For example, the energy-momentum tensor takes the form
\begin{equation}
T_{\mu \nu } = (\alpha \,{\delta ^\lambda }_\mu {\delta ^\rho
}_\nu + \beta \,g^{\lambda \rho}g_{\mu \nu})\psi _{\lambda \rho}
 \, \,, \label{eq17}
\end{equation}
with $\alpha $ and $\beta$ real scalars, and $\psi _{\lambda
\rho}$ be a symmetric tensor containing information about matter
fields. Tensor (17) can describe some interesting systems. If
$\alpha = -1$, $\beta = \frac12$ and $\psi _{\lambda \rho}=
\partial _{\lambda}\phi \partial _{\rho}\phi$ we are talking about
a massless real scalar field $\phi$. On the other hand, if $\alpha
= p+\rho$, $\beta = p$, and $\psi _{\lambda \rho}= U_{\lambda}
U_{\rho}$, we are considering a perfect fluid with density $\rho$,
pressure $p$ and velocity $U_{\lambda}$. In a similar way, taking
suitable definitions can be included an electro-magnetic field or
a bosonic string. Anyway, we assume that $\psi _{\lambda \rho}$
does not depend on metric or on connection (maybe some situation,
where $\psi _{\lambda \rho}$ depends only on metric, would be
considered, but the essentially physical  results will not be much
different).

So, performing $\delta _A S$ variations, one obtains
\begin{equation}
\delta _A S = \int d^N x \sqrt{-g}\,\big(
 {(E^\sigma )^\lambda }_\alpha - 8\pi G\,( \frac
1{\sqrt{-g}}\,\,\partial _\mu (\sqrt{-g}\,\,{(M^{\lambda \mu
})^{\sigma} }_\alpha ) + {[M^{\mu \lambda  } ,A_\mu -T_\mu
]^{\sigma} }_\alpha ) \big) \delta {(A_\sigma
)^{\alpha}}_{\lambda}
 \, \,, \label{eq18}
\end{equation}
up to a boundary term. When equation of motion arising from
$\delta _A S =0$ is written in a torsionless space-time, it leads
to
\begin{eqnarray}
D_{\nu}\big(R^{\alpha \mu}-8\pi G c_1 g^{\alpha \mu}T -8\pi G
c_2T^{\alpha \mu}+c_4 \lambda g^{\alpha \mu} \big)\nonumber \\-
D^{\mu}\big({R^{\alpha}}_{\nu}-8\pi G c_1
{{\delta}^{\alpha}}_{\nu}T -8\pi G c_3{T^{\alpha}}_{\nu}+c_4
\lambda {{\delta}^{\alpha}}_{\nu} \big)\nonumber \\+\,8\pi G
\big(c_2 {{\delta}^{\alpha}}_{\nu} D_\beta T^{\mu \beta}
-c_3g^{\alpha \mu} D_\beta {T^{\beta}}_{\nu}\big)=0
 \, \,, \label{eq19}
\end{eqnarray}
where we had made use of freedom to introduce cosmological terms
with parameter $c_4$. Equation (19) says that $EG \lambda$ is
still trivial solution (our consistence condition), if parameters
take the following values
\begin{equation}
c_4=2c_1=\frac2{N-2}\,\,\,,\,\,\,c_2=c_3=-1
 \, \,. \label{eq20}
\end{equation}

But, as we will show, there are more restrictions on material
fields, this time when {\it n-bein} variations are performed on
action $S$. Obviously, we need to say something about the shape of
$\ell(e,\psi)$. On one hand, this lagrangian density must be
consistent with vacuum limit of the theory ($\ell(e,\psi)
\rightarrow 0$ if $\alpha $ and $ \beta $ go to zero). On the
other hand, we demand consistency with a {\it limit of no
gravitational coupling}, which consists of performing the limit $G
\rightarrow 0$ and $g_{\mu \nu}\rightarrow \eta_{\mu \nu}$ with
$\lambda = 0$. Under these conditions, action (13) must correspond
to the free flat theory of material fields. With this in mind, we
propose a general form for material contribution:
\begin{equation}
\ell(e,\psi)\equiv  L(\psi) +b_1 T^2 + b_2 T_{\mu \nu}T^{\mu
\nu}=q+k\,\psi +b\,{\psi}^2
 \, \,, \label{eq21}
\end{equation}
where $L(\psi)=k_{(\alpha , \beta)}\psi +q_{(\alpha , \beta)}$ is
the lagrangian density of considered matter fields in a curved
background and $b=b_1 (\alpha + N\beta )^2 + b_2 (\alpha
^2+2\alpha \beta +N\beta ^2)$ is a real parameter. Later we study
the limit of no gravitational coupling for parameter $b$.

Then, variations of {\it n-bein} in action (13) can be written in
terms of Ricci, Weyl and material tensors. So, the field equation
is
\begin{equation}
{P^\sigma }_d \big[\psi _{\alpha \beta}, {e^\mu}_b ,R_{\mu
\nu}\big]+{Q^\sigma }_d \big[ {e^\mu}_b ,R_{\mu
\nu}\big]+{S^\sigma }_d \big[\psi _{\alpha \beta}, {e^\mu}_b
,R_{\mu \nu},C_{\mu \nu \alpha \beta}\big]=0
 \, \,, \label{eq22}
\end{equation}
where ${P^\sigma }_d$ and ${Q^\sigma }_d $ are quadratical
polynomial functionals on $\psi _{\alpha \beta}$ and Ricci tensor,
respectively. Moreover,
\begin{eqnarray}
S_{\alpha \beta} \equiv  C_{\mu \nu  \lambda \alpha}{C^{\mu \nu
\lambda }}_{\beta}-\frac{g_{\alpha \beta}}{4} C_{\mu \nu \rho
\lambda }C^{\mu \nu \rho \lambda }-C_{\mu \nu \lambda \alpha
}{R^{\mu \nu \lambda }}_{\beta }-C_{\mu \nu \lambda \beta }{R^{\mu
\nu \lambda }}_{\alpha }\nonumber \\+ \frac{g_{\alpha \beta}}{2}
C_{\mu \nu \rho \lambda }R^{\mu \nu \rho \lambda }-16\pi G\,\alpha
\, C_{\alpha \lambda \beta \rho}{\psi}^{\lambda \rho}\, \,,
\label{eq23}
\end{eqnarray}
where $C_{\mu \nu  \lambda \alpha}$ is the Weyl tensor. At this
point, one can explore the particular case in which $N=3$ because
the Weyl tensor is null identically. Then, if one expect that
action (13) must be an extremal on  $EG \lambda$ we evaluate
equation of motion for dreibein (22)
\begin{eqnarray}
(8\pi G)^2\big(-\frac{4b}{(8\pi G)^2}+4{\alpha }^2 -26\alpha \beta +2{\beta}^2\big)\psi {\psi ^\sigma}_d \nonumber \\
+(8\pi G)^2\big(\frac{b}{(8\pi G)^2}-{\alpha }^2 -4\alpha \beta -7{\beta}^2\big){\psi}^2{e^\sigma}_d    \nonumber \\
+8\pi G\big((22\alpha -12\beta)\lambda -16\pi
G \alpha a -\frac{k}{4\pi G}\big){{\psi}^\sigma}_d \nonumber \\
+8\pi G\big((2\alpha -8\beta)\lambda -16\pi G \beta a
-\frac{k}{8\pi G}\big)\psi
{e^\sigma}_d \nonumber \\
+(q+16\pi G a\lambda ){e^\sigma}_d =0 \, \,. \label{eq24}
\end{eqnarray}

The last equation of motion necessarily represents a restriction
for material fields. Therefore the following restriction arise
\begin{equation}
\psi =constant \,\,\epsilon \,\,R_e \, \,, \label{eq25}
\end{equation}
 on possible field
configurations. This is not severe in case of a perfect fluid
(i.e., $\psi = U^\mu U_\mu =-1$). Anyway, for all ${\psi
^\sigma}_d$ with $\psi =constant $, equation (24) gives two
relations for $a$ and $b$
\begin{equation}
2 \psi b + (8\pi G)^2 \alpha a =(8\pi G)^2(2{\alpha }^2 -13\alpha
\beta +{\beta}^2)\psi + 8\pi G(11\alpha -6\beta )\lambda -k \, \,,
\label{eq26}
\end{equation}
\begin{equation}
 \psi ^2 b +  16\pi G(\lambda-8\pi G \beta \psi )a =(8\pi
G)^2(\alpha ^2 +4\alpha \beta +7\beta ^2)\psi ^2 -8\pi G(2\alpha -
8\beta ) \lambda \psi -k\psi -q\, \,. \label{eq27}
\end{equation}
In order to obtain regular solutions one need to demand the new
restriction
\begin{equation}
\psi (4\lambda - 8\pi G(\alpha +4\beta )\psi ) \neq 0 \, \,.
\label{eq28}
\end{equation}

With the last condition one can study the limit at no
gravitational coupling for $b$
\begin{eqnarray}
b\mid _{\lambda =0}=(8\pi G)^2\bigg(\frac{\alpha ^3 +8{\alpha }^2
\beta-26\alpha \beta ^2 +9\beta ^3}{\alpha +4\beta}
\bigg)-\frac{(\alpha
 +2 \beta )k\psi +q\alpha}{(\alpha
+4\beta)\psi ^2}\, \, ,\label{eq29}
\end{eqnarray}
being consistent in case of a massless scalar field, in which
$\alpha =-1$, $\beta =1/2$ and $q=0$ (i.e., when $G\longrightarrow
0$, relation (29) gives $b\mid _{\lambda =0} =(8\pi G)^2
(...)\longrightarrow 0$). This is not the same situation for a
fluid, because the limit $b\mid _{\lambda =0} \longrightarrow 0$
demands an additional restriction on density and pressure:
$(p+\rho)p=0$ (i.e., one should consider dust).

 We want to underline that eq.(25) and eq.(28)
means that non minimal coupling scheme presented in eq.(13) is
consistent with Einstein gravity only under certain conditions
related with the class of material distribution and the features
of space-time (i.e., restrictions on possible values of
cosmological constant). This is not a surprising idea. In fact,
from the point of view of high spin gauge fields coupled to
gravity it can be found that theory is consistent only in
restricted backgrounds$^{12}$. So, maybe new interaction  terms
(i.e., involving auxiliary fields ) added to action (13) enclose
the hope to reduce the number of constraints on material fields.

In this sense, let $S'$ be the new action with $J^{a}$ and
$H^{\alpha \beta}$ some functional on dreibein and material
fields, and ${(W_a)^{\mu}}_{\nu}$ the components of a auxiliary
field, which transforms like a $GL(3,R)$ connection, then we write
\begin{eqnarray}
S'=S_o + \int d^3 x \sqrt{-g}\big(\ell(e,\psi) + 4\pi
G\,tr\,M^{ab}(e,\psi)F_{ab}  \nonumber \\
+\,tr\,J^{a}(A_a-W_a)+tr\,H^{a b}(A_a-W_a)(A_b-W_b)\big)
 \,\, \,, \label{eq30}
\end{eqnarray}
where a naive proposal for components of $J^{a}$ and $H^{a b}$ is
taken
\begin{equation}
(J_{\beta})_{\mu \nu }\equiv (d_1 + d_2 \psi){\varepsilon }_{\beta
\mu \nu} \, \,, \label{eq31}
\end{equation}
\begin{equation}
( H^{\alpha \beta})^{\mu \nu} \equiv a_1
\,g^{\alpha\beta}g^{\mu\nu}+a_2 \,g^{\alpha\mu}g^{\beta\nu}+ a_3
\,g^{\alpha\nu}g^{\beta\mu}
 \, \,, \label{eq32}
\end{equation}
with the real parameters $d_1$, $d_2$, $a_1$, $a_2$ and $a_3$.

 From eq.(30), equation of motion for $W_b$ is
\begin{equation}
J^{b} + H^{a b}(A_a-W_a)+(A_a-W_a) H^{ba}=0
 \, \, ,\label{eq33}
\end{equation}
establishing that field equations of connection fields $A_a$ are
maintained with or without introducing auxiliary terms. On the
other hand, suggests an ansatz for auxiliary fields:
\begin{equation}
(A_{\alpha}-W_{\alpha})_{\mu \nu}= (\theta _1 +  \theta _2 \psi )
\,{\varepsilon }_{\alpha\mu\nu}\, \,, \label{eq34}
\end{equation}
with $\theta _1$ and $\theta _2$ real parameters (this is not the
most general linear dependence on field $\psi _{\sigma \beta}$,
but it is sufficient and consistent with eq.(33)). Ansatz (34)
give the relations $d_1 = 2(a_2-a_1)\theta _1$ and $d_2 =
2(a_2-a_1)\theta _2$.

Next, equation of motion for dreibein is evaluated on $EG \lambda
$
\begin{eqnarray}
\bigg({P^\sigma }_d \big[\psi _{\alpha \beta}, {e^\mu}_b ,R_{\mu
\nu}\big]+{Q^\sigma }_d \big[ {e^\mu}_b ,R_{\mu \nu}\big] + \,tr
\frac{\delta J^{\beta}}{\delta
{e_\sigma}^d}(A_{\beta}-W_{\beta})\nonumber
\\+\,tr \frac{\delta H^{\alpha \beta}}{\delta
{e_\sigma}^d}(A_{\alpha}-W_{\alpha})(A_{\beta}-W_{\beta}) -tr
H^{\alpha \beta}(A_{\alpha}-W_{\alpha})(A_{\beta}-W_{\beta})
\,\,{e^\sigma}_d\bigg)_{EG\lambda }=0\, \,. \label{eq35}
\end{eqnarray}

Then, using eq.(31), eq.(32) and eq.(34), it can be found for any
material field configuration an equation system for free
parameters
\begin{equation}
16\pi G a \lambda -14(a_1 -  a_2){\theta _1}^2=-q_{(\alpha , \beta
)} \,, \label{eq36}
\end{equation}
\begin{equation}
 2(8\pi G)^2\beta a +12(a_1 -
a_2)\theta _1\theta _2=8\pi G(2\alpha -8\beta)\lambda +k_{(\alpha
, \beta )} \,, \label{eq37}
\end{equation}
\begin{equation}
(8\pi G)^2\alpha a -12(a_1 -  a_2)\theta _1 \theta _2=8\pi
G(11\alpha -6\beta)\lambda -k_{(\alpha , \beta )}  \,,
\label{eq38}
\end{equation}
\begin{equation}
b-14(a_1 - a_2){\theta _2}^2=(8\pi G)^2(\alpha ^2 +4\alpha \beta +
7\beta ^2) \,, \label{eq39}
\end{equation}
\begin{equation}
2b-12(a_1 - a_2){\theta _2}^2=(8\pi G)^2(2\alpha ^2 -13\alpha
\beta + \beta ^2) \,. \label{eq40}
\end{equation}

Consistence with limit at no gravitational coupling for $b$ can be
verified and, furthermore only for free parameters $a$, $b$, $(a_1
- a_2){\theta _1}^2$, $(a_1 -  a_2){\theta _2}^2$ and $\lambda$,
this system can be solved. From this, just restrictions on
possible values for cosmological constant remains.

We finalize, saying that although the coupling terms presented in
eq.(30) (with eq. (31) and (32)) do not have the most general
form, the idea that strong restrictions on ${\psi }_{\alpha
\beta}$ can be avoided with auxiliary fields has been elucidated.
A possible extension of this type of study pointing to
metric-affine gravity, including
 non-Riemannian interaction terms$^9$, must be performed elsewhere.
\vskip .2truein
 {\bf Acknowlegdments}

 Author wish to thanks Prof. P.J. Arias
 for discussions and Prof. F. Mansouri for references. And also
 the referee for the remarks. This work is supported by Alma
 Mater-OPSU program.

\end{document}